\documentclass[a4paper,11pt]{article}
\usepackage{pos}
 
\def\A{{\cal A}}
\def\B{{\cal B}}
\def\Tr{{\rm Tr}}
 \newcommand{\SO}{\mathrm{SO}}
 \newcommand{\SU}{\mathrm{SU}}
 \newcommand{\unit}{\mathrm{U}}
 \newcommand{\aat}[4]
{\tilde{A}^{#1}_{#2}\atop(({#3}),({#4}))}
\usepackage{adjustbox}
\def\A{{\cal A}}
\def\B{{\cal B}}
\def\Tr{{\rm Tr}}

\usepackage{adjustbox}

\begin{document}
\title{Domain Walls of  Low Tension in Cosmology }
 \ShortTitle{Low tension walls}
\author*[a]{Holger Bech Nielsen}
\author[b]{Colin D. Froggatt}
\affiliation[a]{Niels Bohr Institute,\\ Blegdamsvej 15 -21, Copenhagen, Denmark}
\affiliation[b]{School of Physics and Astronomy, Glasgow University,\\
  Glasgow G12 8QQ, Scotland}
\emailAdd{hbech@nbi.dk}
\emailAdd{Colin.Froggatt@Glasgow.ac.uk}
\abstract{In the present article we put up for discussion the idea of there
  existing several versions/phases of the vacuum, in the spirit in which
  we have long worked on this idea, namely the Multiple Point Citicality
  Principle, which also says that these different vacuum phases have the
  same energy density. 
  We mention that we indeed predicted the Higgs mass to be ($135\pm 10$) GeV, which when measured turned out to be 125 GeV, using the
  assumption of this Multiple Point Criticality Principle. 
  We consider the possibility that there is one type of vacuum in the galaxy clusters (the usual vacuum) and another type of vacuum in the voids. The hope that there could indeed be such a low tension $S$ of the domain wall between
  these two phases, that it would not totally upset cosmology is based on our dark matter model. In this model 
 %fit to a model of ours in which the 
 dark matter consists of pearls or bubbles
  of a new vacuum phase, with ordinary matter inside it under very high
  pressure. The
  order of magnitude of $S^{1/3} \sim MeV \; or  \; 100\,  MeV $ could make such
  domain walls
  astronomically viable. We successfully estimate the order of magnitude of the
  variations in the fine structure constant in different places astronomically,
  but the similar variations in proton mass over electron mass should have been
  much bigger than seen experimentally in our model. The Universe's  surprisingly early galaxies seen by
  JWST (James Webb telescope) may agree well with our model. Replacing the usual cosmological constant by domain walls in the standard cosmological model would lead to a cubic root of the tension being $S^{1/3}\sim 30\, MeV$. 
   }
\FullConference{CORFU2022: 22th Hellenic School and Workshops on Elementary
  Particle Physics and Gravity\\ 
 Workshop on Tensions in Cosmology\\
Corfu\\
Sep 07 - Sep 12, 2022
}
%% \tableofcontents
\maketitle

\author{H.B. Nielsen
  %\footnote{Speaker at the  Work Shop
  %``What comes beyond the Standard Models'' in Bled.}
  , Niels Bohr Institut,
C.D.Froggatt, Glasgow University}
\date{Korfu 9. of  September  2022}

%\begin{document}
\section{Introduction}
\begin{figure}        
  \includegraphics[scale=0.8]{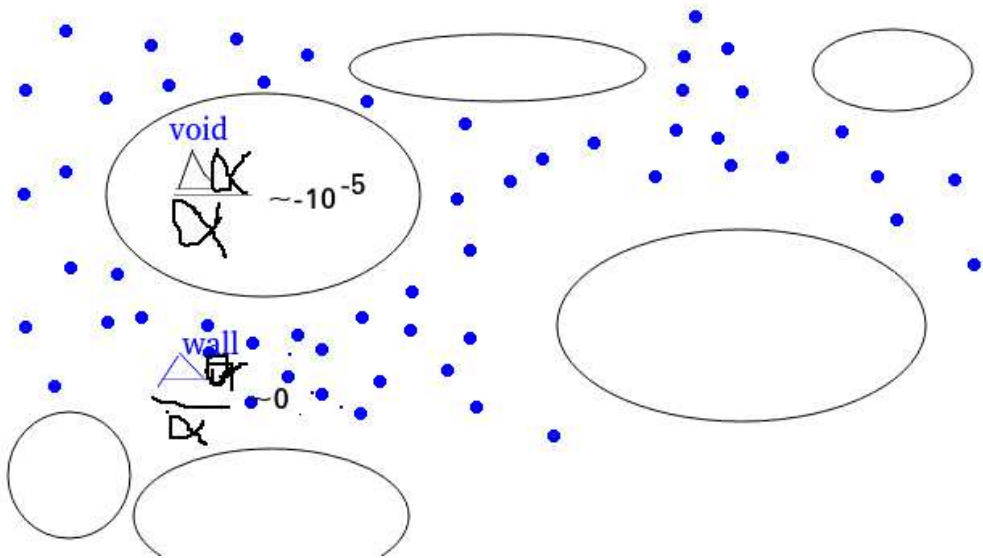}
  \caption{\label{voids}
  {\em ``Artistic impression'': Vacua in Universe; Voids one vacuum, clusters of
    galaxies ``our'' vacuum.} The small dots symbolise galaxies of course.
  The formulas should illustrate that we imagine that the fine structure
  constant $\alpha$ should deviate a little bit $\Delta \alpha $ from the usual
  value, or say the value on Earth, when one measures it in a void. The formula
  suggests that the deviation is relatively of the order of $10^{-5}$ in the
  voids, while assuming the Earth to be in a galaxy dense region there should be
  no deviation between the Earth-value and the galaxy rich region value.}
  \end{figure}

%\section{Overskrift}
%\begin{frame}
%  \titlepage
%\end{frame}
%
%\begin{frame}
Investigating the three dimensional distribution of the galaxies shows that
the galaxies cluster into surfaces and line-like structures, so as to leave
a lot of voids with relatively much fewer galaxies. The main speculation
of the present contribution is that the idea of there being several versions
of the vacuum, which we used in the talk \cite{Corfu2022} in the foregoing workshop
``Standard Model and Beyond'' as the speculative input in our model for
dark matter\cite{Dark1, Dark2, Tunguska, supernova, Corfu2017, Corfu2019, Bled20, Bled21, theline, extension, Corfu21}, could
also mean that there were different vacua in different
regions in outer space. For example, one could think that there were one type of
vacuum in the voids and another type in the galaxy rich regions.
In our model of dark matter,
a different vacuum from the usual one keeps some ordinary matter under high
pressure in a very concentrated bubble, so that it is essentially hidden
%from being ``seen'' 
and does not interact much with ordinary matter despite the high mass
of the bubble. Thus we have a picture of the dark matter as genuinely containing some ordinary matter in it; in this way it has a slight similarity
to Khlopov's model \cite{Khlopov, Khlopov1, Khlopov3, Khlopov5} in
which a helium nucleus is part of the dark
matter complex.
In Figure
\ref{voids} the voids in the Universe are marked by ellipses in a picture of the galaxy pattern drawn in perspective. If you take these ellipses as symbols
for the boundaries between the different vacuum phases, then you could consider this figure to suggestively illustrate how there is speculated to be one phase of the vacuum in the voids and another in the galaxy rich regions.

%\section{Low Tension}
%\begin{frame}
  \section{ Domain Walls need Low Tension to be Allowed}

  The tension or in $c=1$ notation also the energy per area $S$ of a domain
  wall between different phases of the vacuum has dimensionality $GeV^3$. If
  an energy scale of the order of the scales at which we look for
  `` new physics'' is used, the energy density gets so high that domain walls
  of cosmological scales of extension would be so heavy as to be totally
  excluded by the already known energy density (the critical density):
  \begin{eqnarray}
    \rho_c &=& \frac{3H^2}{8\pi G}= 1.88*10^{-26}h^2kg/m^{3}\\
    &&\hbox{(where $h=H_0/(100km/Mpc) = 0.67$)}\\
    &=& 2.78*10^{11}h^2M_{\odot}Mpc^{-2},\\
    &=& 8.5*10^{-27}kg/m^3.
    \end{eqnarray}
The average energy density that would be produced by even just a
single domain wall crossing the visible universe with energy density $S$ of
the order of the
``new physics'', meaning say TeV scale to the third power $S\sim TeV^3$,
would dominate the critical energy density in the Universe.
%that our present cosmology would be totally dominated out. 
So we can conclude that {\em  ``New physics scale'' walls are excluded.}

%\end{frame}

%\begin{frame}

 \section{ How high an Energy per Area is possible? \label{section 3}}

Consider a domain wall having a dimension of the visible Universe length scale:
\begin{eqnarray}
	R_{visible} &=& 13 *10^9 \; \hbox{light} \; \hbox{years}\\
	&=& 1.3*10^{10}ly *9.5*10^{15}m/ly = 1.2 *10^{26}m,
\end{eqnarray}
which therefore has 
\begin{eqnarray}  
	Area &\sim & 10^{52}m^2.
\end{eqnarray}
%for a domain wall would allow an energy per area to give 
Now the total visible critical energy is
\begin{eqnarray}
	\hbox{visible (critical) energy} &\sim & (1.2 *10^{26}m)^3
	*8.5*10^{-27}kg/m^3\\
	&=& 1.5*10^{52}kg.
\end{eqnarray}
So this would allow the domain wall to have an energy up to 
\begin{eqnarray}
	\hbox{maximal energy per area } &\sim& \frac{1.5*10^{52}kg}{10^{52}m^2}
	\sim  1.5 kg/m^2.
\end{eqnarray}
%\end{frame}
%\begin{frame}
Hence the tolerable energy per area for a domain wall of astronomical extension is $\sim 1.5 kg/m^2$:
\begin{eqnarray}
	\hbox{Tolerable energy per area = }S &=& 1.5 kg/m^2\\
	&=& \frac{1.5 kg/m^2*(1.2*10^{-12}MeVm)^2}{1.79*10^{-30} kgc^2/MeV}\nonumber\\
	&=& 1*10^{6}\, MeV^3.
\end{eqnarray}

So the energy scale pointed out by such a slightly too big domain
wall energy per area is
\begin{eqnarray}
	\sqrt[3]{S} &=& \sqrt[3]{10^6 MeV^3} = 100\, MeV.
\end{eqnarray}

  %\end{frame}
   \begin{figure}
   	\includegraphics[scale=0.55]{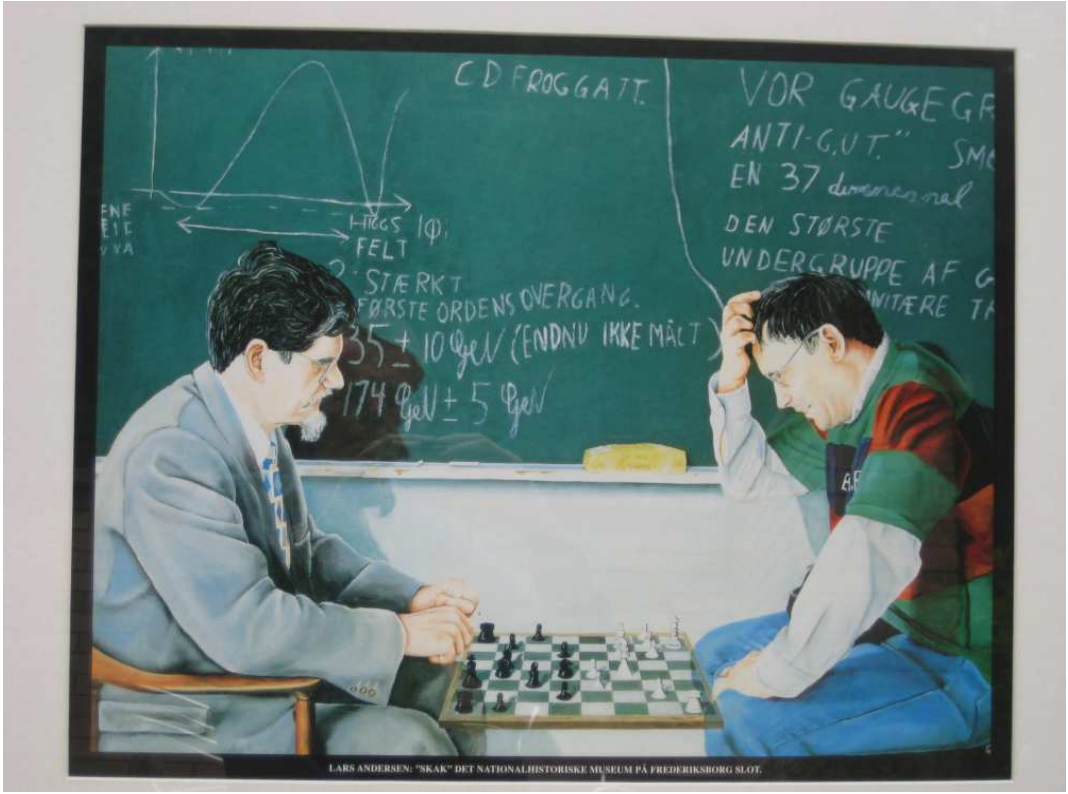}
   	\caption{\label{Lykketoft} This painting was put up at Frederiksborg Castle
   		near Hiller{\o}d {\em before} the Higgs particle was found in 2012, so nobody knew
   		the mass of the Higgs then. Nevertheless we had already published the
   		predicted Higgs mass $135 \pm 10$ GeV and except for the $1$ hidden by
   		Mogens Lykketofts head you see this number on the painting.}  
   \end{figure}
   
   \section{Older Achievements}
   % \begin{frame}
   
   We have worked for a long time on the idea of there being several vacua with the same
   energy density and this has already led to some achievements:
   
   \begin{itemize}
   	\item {\em PRE}dicted the Higgs mass {\em before} it was found \cite{tophiggs, Corfu1995}. See Figure \ref{Lykketoft}.
   	
   	\item In a rather complicated model with each family of fermions
   	having its own set of gauge fields - like the ones in the Standard Model - one of us (HBN), Niels Brene, Don Bennett, I Picek and others \cite{Picek,ActaPP, GG} fitted the fine structure
   	constants by using the
   	number of families as a parameter and {\em got 3 before this number was
   		measured.}
   	\item We model dark matter
   	particles or pearls \cite{extension, Dark1, Dark2, Tunguska,
   		Corfu2017, Corfu2019, Corfu21, Bled20, Bled21, theline, supernova} as bubbles of a new vacuum containing highly
   	compressed ordinary matter. We get the surface tension $S$ of the domain
   	wall between the vacua to have its cubic root of the order
   	\begin{eqnarray}
   		S^{1/3} &\sim& 10\, MeV.
   	\end{eqnarray}
   An important part of our fitting in this model involves the 3.5 keV X-ray line observed from satellites \cite{Bulbul, Boyarsky, Boyarsky2}.
   \end{itemize}
   %19\%

  %  \end{frame}
\section{Big Bubbles}
%\begin{frame}
The main point of the present talk is to point out that, once you have
the suggestion that the tensions $S$ of the domain walls  between different
vacuum-versions could
be as small as the numbers mentioned $(10\, MeV)^3$ or $(100\, MeV)^3$, then
astronomically extended domain walls become a possibility. One concrete
idea could be that the domain walls lie around the voids in outer space
containing a very low density of galaxies. That is to say we propose the  
{\em Idea of Domain Walls around Voids}:

Since our dark matter model has one vacuum inside the dark matter, a
large region of such inside the dark matter vacuum would be formally
a huge pearl of dark matter, and there could not be any true dark matter
inside it.
%Thus it would be much less easy for the contraction to get
%started early in such a region.
Such a difference would almost certainly have an influence on the number
of galaxies formed at different times. But thinking about the likely
situation inside such a huge bubble with the inside the dark matter
vacuum, we should imagine that the usual dark matter there has been replaced
by an equal amount (by mass) of ordinary matter. In such a situation
there would be a better chance for the material when contracting to
emit radiation and cool down, than in a usual region with both
dark matter and (much less) ordinary matter. Thus we would at first
expect a faster formation of galaxies and stars in such a region of
the ``inside dark matter vacuum'' than in a region like our own.
The matter will namely fall faster together. In turn this could mean that
the faster start means a faster burn out and that the galaxies in the
``inside dark matter vacuum'' regions would 13 milliard years later, in our epoch, send out less light than the usual regions.
So today it could be that the galaxies would look sparser, because of being
older and giving less light.

In  the regions near the domain wall - the outermost part of the
``inside dark matter vacuum'' region -  where the nuclei would probably
come in from outside and get speeded up to high temperature,
there would be a high temperature and at least in the beginning not much
star formation.

%If further as we shall see a region
%of inside-dark matter-vacuum would tend to be heated up, then the chanse for
%forming stars and galaxies would also later be reduced.

\begin{figure}
    \includegraphics[scale=0.5]{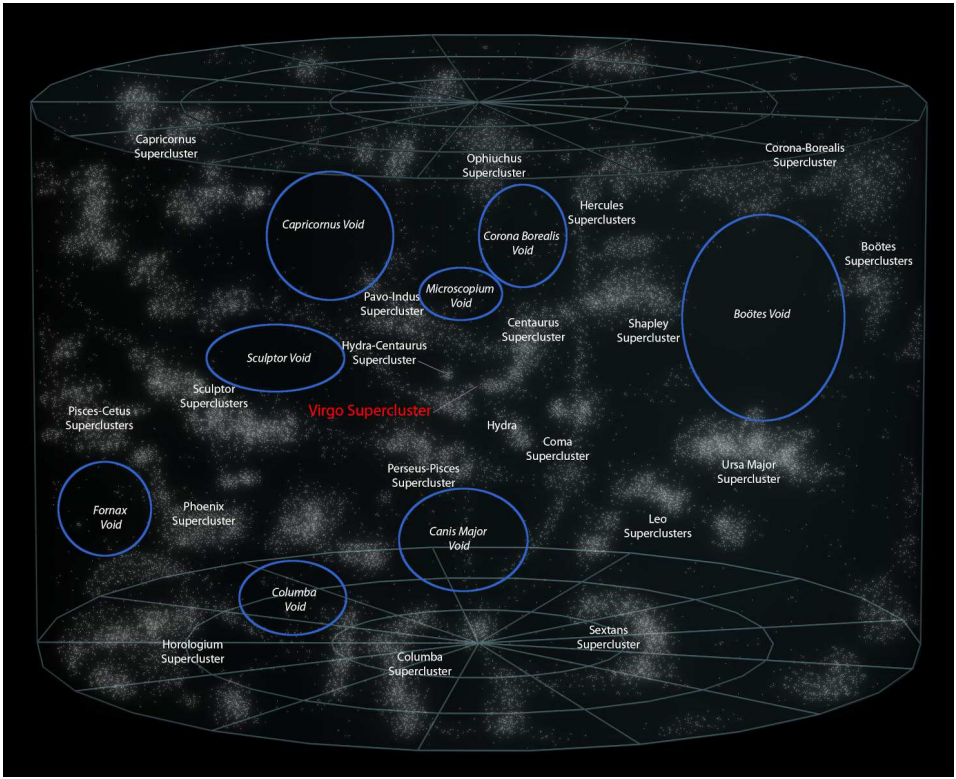}
    \caption{\label{voids2} Here is shown in perspective the measured galaxy
      densities and the voids with very few galaxies are marked with rings and
      names. Accidentally these rings just made (in Wikipedia) by the author of the figure
      to point out the voids could be considered our proposal for where to
      find the domain walls.} 
    \end{figure}

So let us investigate as an example the hypothesis that the domain walls
follow the borders of the voids - the regions with few or no galaxies -
%because
so that these voids really are ``formally huge dark matter pearls''. This idea is illustrated in Figure \ref*{voids2}

  % \vspace{-3mm}

\section{MPP}
%\begin{frame}

We originally proposed our `` Multiple Point Criticality Principle'' (= MPP) \cite{MPP1,MPP2,MPP3,MPP4}
to justify that we fitted e.g. the fine structure constants in a somewhat complicated model with the phase transition values taken from a lattice model. That is to say we would
generally argue that there should be in Nature some law saying that the values of coupling constants should be made to sit on the phase transition point(s).

\begin{figure}                
	\includegraphics[scale=1.2]{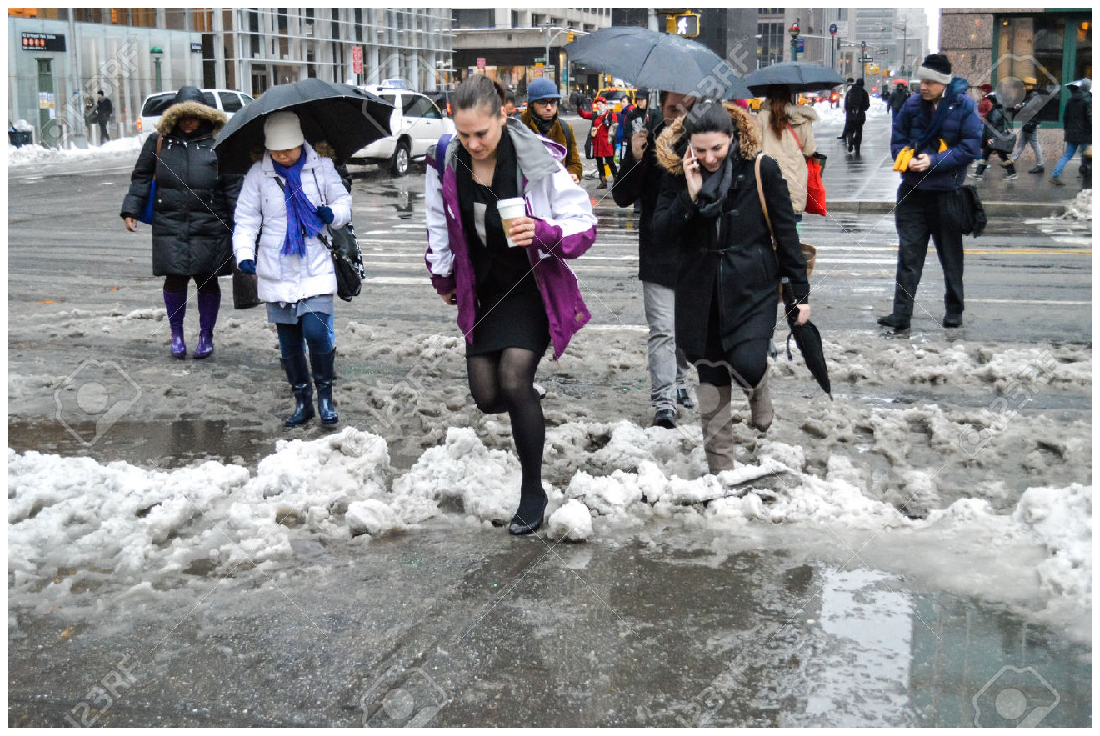}
	\caption{Slush $\Rightarrow$ $0^0C$. }
	\label{slush}
\end{figure}

This is what (often) happens with a microcanonical system - meaning a system
in which the energy is fixed - when it splits up and you get more than one
phase at the same time so as to achieve the given total energy. For example, you may
get as in Figure \ref{slush} both ice and fluid water together provided
there is a (fixed) given amount of energy, as in the microcanonical ensemble.
It should be stressed that even though the fixed energy is {\em not fine-tuned}
you end up with more than one phase existing together, and that can only happen
in the case of slush say at one specific (fine tuned) temperature $0^0C$.
So one got the fine tuning without putting any fine-tuning in! Most often the
two or several phases achieved in this way, by fixing {\em extensive quantities} (like total energy or the total amount of some element or some molecule) will occur
(at the end) in comparable amounts.

The suggestion is that it is a similar mechanism to the microcanonical ensemble,
that causes the coupling constants in Nature to tend to be precisely at
phase transition points - the ``multiple point''. But then, by analogy
to the slush etc., there should be extended amounts of the different phases of {\em comparable size}. It is these extended phases, which we think of as being different
versions of the vacuum. The proposed mechanism for why we should have these
different phases of the vacuum predicts that they should occur
in order of magnitudewise comparable amounts. So it would not be expected
that the one phase would only occur inside some tiny dark matter bubbles, while
another phase would be extended over the whole world and fill almost everything.
No, with this idea of extensive quantities fixed, the analogy suggests
that also the inside the dark matter bubbles phase must occur as astronomically
extended huge bubbles, e.g. in the voids.

%\end{frame}
%\begin{frame}

%                {\bf Several degenerate Vacua (MPP) $\sim$ Slush %$\Rightarrow$ $0^0C$}
%\end{frame}

\section{Vary $\alpha$}
%\begin{frame}
We consider here, as a signal for the above suggested existence of several vacuum phases in comparable amounts, the possibility that the natural constants
%We hope for obtaining as a major signal for, that there indeed should
%such several vacuum phases as we have suggested from the analogy to the
%microcanonical ensample, which suggests that in the generic situation there
%shall be comparable amounts of the different phases, that the natural %constants
such as the fine structure constant $\alpha$ should be different in the
different phases. As it is possible to observe the fine structure
constant in different clouds of atoms of hydrogen, such an investigation has
indeed been performed although the accuracy has so far only been barely
sufficient to give convincing results.

\subsection{Expected Deviations from Constant $\alpha$}

But let us first estimate the order of magnitude of the deviation from a constant
fine structure $\alpha$ from vacuum phase to vacuum phase, which we would
predict using the parameters extracted from our long studied model
for the dark matter:

  %{\bf Parameters of Domain Walls from Our Dark Matter Fitting}
  \begin{itemize}
  \item Cubic root of tension in wall $S$ or of energy density per
    unit area
    \begin{eqnarray}
      \sqrt[3]{S} &=& 10\, MeV \quad \hbox{ (allowed for cosmological domain
        walls)}.\nonumber
    \end{eqnarray}
    We extracted this tension from the assumption that we could estimate
    how large the typical size of our bubbles making up the dark matter would
    be and the pressure inside the bubbles. This pressure was estimated
    from the frequency of the radiation of X-rays emitted from the dark matter
    pearls being identified with the mysterious 3.5 keV radiation line
    observed in X-ray spectra \cite{Bulbul, Boyarsky, Boyarsky2}.

  \item  Lowering of the potential energy for a nucleon by entering into the
    ``new'' vacuum (the one inside the dark matter bubbles)
    \begin{eqnarray}
      \Delta V &=& 2\, MeV. 
    \end{eqnarray}
    Nucleons 2 per mille lighter in the ``new'' vacuum.

     Actually we found \cite{Bled21} that both the frequency of the 3.5 keV X-ray line and the overall magnitude of this radiation as obtained by
    Cline and Frey \cite{Cline} could be fitted with just one
    %(with a few observations not agreeing though) 
    parameter combination $\frac{\Delta V}{\xi^{1/4}}$. Here $\xi$ is the ratio
    of the typical dark matter bubble radius to the critical one below which
    the nuclei in it get spit out. This was in a model, wherein the X ray emission
    came from colliding dark matter pearls uniting with each other.
    
  \item Energy density in the two vacua exactly the same. (Also a theorem
    by Gia Dvali).

    This was the original main assumption of the 
    ``Multiple Point Criticality Principle'' (=MPP), which we mentioned gave
    us the Higgs mass prediction. But it was actually proven by Gia
    Dvali \cite{Dvali} that it should not be possible to have several vacuum
    phases unless they have the same energy density.
    \end{itemize}
 % \end{frame}

  \subsection{Calculation of the Order of Magnitude of the Difference in $\alpha$
    from phase to phase.}
%\begin{frame}
  A physical argument for the change of the (effective) fine structure
  constant:

  Because of self energy diagram corrections, we should think of the
  photon as being replaced with some amplitude by a proton anti-proton
  pair virtually. The probability for finding the photon in this virtual
  state of a proton anti-proton pair is proportional to the fine structure
  constant, because the coupling causing the transition is proportional to
  $e$ the charge quantum and the probability for finding the proton
  anti-proton pair will go as the square of this transition (of course
  $\alpha \propto e^2$). But it is also inversely proportional to the excess
  energy of the proton anti-proton state; this is the well-known second
  order perturbation formula for the energy shift in quantum mechanics.

  The effect of such a partial virtual replacement of the bare photon
  by the proton anti-proton pair is to diminish the coupling of the
  bare photon, because it is part of the ``time'' (really probability)
  replaced by the not directly coupling pair. The bigger the chance of
  finding the pair the smaller the coupling of the dressed photon propagator.
  This probability for finding the pair, instead of the bare photon itself,
  increases with lower energy or mass of the pair. So the coupling
  $\alpha$ is the smaller in the case of the lighter proton and anti-proton.
  This lighter pair situation is in our model for dark matter supposed to
  happen in the inside the dark matter phase. So we predict that the smaller
  fine structure constant should be found in the inside the dark matter phase.
  But since we have dark matter around in our galaxy, we must exist in a
  phase of outside the dark matter vacuum. So we should ourselves be in a
  phase with slightly bigger $\alpha$. Thus we predict that the deviating
  $\alpha$ clouds should have a lower $\alpha$ than the value here.

  The order of magnitude is expected to be proportional to the ratio
  $\frac{\Delta V}{m_p}$ where $m_p$ is the proton mass and $\Delta V$ is
  our estimated relative potential energy for a nucleon in the
  two phases, which we claim to be of the order of $\Delta V \approx 2 MeV$.
  
  Let us now consider the proton loop self energy diagram for the photon in the two phases: 
  %as shown in Figure \ref{diagram}. 

%  {\bf E.G. Photon self energy diagram Proton loop part gets changed}
%\begin{figure}
%  \includegraphics[scale=0.6]{photonself4.eps}
%  \caption{\label{diagram} Diagram estimating the effect on the
%    fine structure constant $\alpha$ from  going into a phase with
%    a different proton mass. }
%  \end{figure}
 % \end{frame}

\begin{figure}
  	\includegraphics{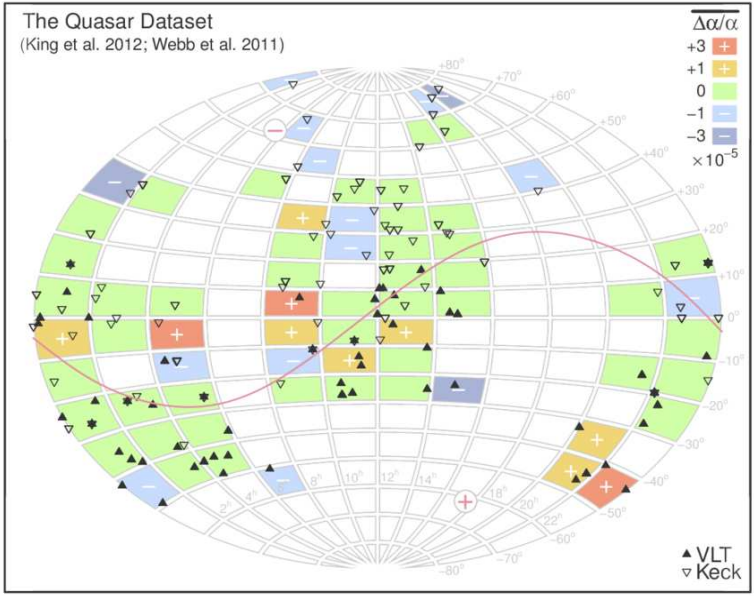}
  	\caption{\label{sky-graph} A sky graph \cite{variation} giving the small regions in which
  		the various deviations for the fine structure constant have been measured,
  		because appropriate clouds of atomic gas had been radiated from behind by quasars and the generated absorption lines observed.
  		The agreement with our prediction of the sign,
  		that $\frac{\Delta \alpha}{\alpha} \le 0$ is really not well satisfied,
  		in  as far as it should mean on this plot that there were no red nor
  		yellow squares, but only the no deviation green squares and the blue
  		ones which have positive deviations. But the order of magnitude is
  		o.k.
  		Strictly speaking we would predict only two different values, but
  		clearly there are huge statistical fluctuations (indeed the uncertainty
  		would allow for
  		no deviations at all).}
  \end{figure}

%\begin{frame}
 % {\bf Sign and order of magnitude of $\frac{\Delta \alpha}{\alpha}$}
  \begin{itemize}
%\vspace{-4mm}    
  \item In the approximation that one could use non-relativistic second order
  perturbation theory for the sign, we expect that the (bare) photon is
  mixed more with the proton anti-proton pair in the phase with the smallest
  nucleon masses. Thus the photon in this ($m_p$ small) phase couples 
  weaker than in the $m_p$ big phase.

\item The relative correction to the fine structure constant $\alpha$, i.e.
  $\frac{\Delta \alpha}{\alpha}$, has one factor $\alpha$ in addition
  to the $2/940$, the relative difference in nucleon mass between the two
  phases:
  \begin{eqnarray}
    \frac{\Delta \alpha}{\alpha} & \approx & - \alpha *\frac{2}{940} =
    -1.4*10^{-5}
  \end{eqnarray}
\item The deviations found (see Figure \ref{sky-graph}) are of this $10^{-5}$
  order of magnitude, but in the literature they are fitted by
  a dipole variation in space \cite{dipole1,dipole2} rather than with voids versus
  clusters.
  \end{itemize}
  
 % \section{A Result Sky-graph}
 More detailed observations are needed before a proper check of our model can be made, but in Figure \ref{sky-graph} we show a sky-graph from reference \cite{variation} with the results found so far. It is necessary to 
 investigate whether the clouds studied were in voids or in galaxy rich regions. We predict that deviations $\frac{\Delta \alpha}{\alpha}$ shall be of order $10^{-5}$ giving smaller values of $\alpha$ in the voids than in the galaxy rich regions. But whether you have a
  void or a galaxy rich region not only depends on the direction but also
  on the distance, which in practice has to be taken as the red shift $z$.

 % \end{frame}
%\begin{frame}
 \section{$m_p/m_e$ ratio not varying at $10^{-6}$ level is problematic for our model}

Searches have also been made to detect cosmological variations in the proton to electron mass ratio $\mu=\frac{m_p}{m_e}$ using molecular absorption lines and $\mu$ was found not to vary by more than one part in a million.
  %Using molecule absorption lines the ratio of the electron to
  %proton mass was found not to vary by more than one part in a million.
  
  The wavelengths of the Lyman and Werner transitions of the $H_2$ molecule are sensitive to $\mu$ and King et al \cite{King} used the spectra from 3 quasars, viewed through molecular clouds, to limit the variation in $\mu$ relative to the present day laboratory measured value to $\frac{\Delta \mu}{\mu} = (2.6 \pm 3.0) \times 10^{-6}$. Also the inversion transitions of ammonia are particularly sensitive to $\mu$ when compared to molecular rotational transitions. Murphy et al \cite{Murphy} used this technique to limit relative deviations from the laboratory value to $\frac{\Delta \mu}{\mu} < 1.8 * 10^{-6}$. Later Kanekar \cite{Kanekar} improved this limit  to obtain $\frac{\Delta \mu}{\mu} < 3.6 \times 10^{-7}$. Methanol transitions were then used \cite{Bagdonaite} to find the more stringent limit  $\frac{\Delta \mu}{\mu} < 1.8 * 10^{-7}$.  
  
%  T.D. Le \cite{metomp} even finds
%  $\frac{(\Delta \mu}{\mu} =-0.360 \pm 0.864)\times 10^{-8}$
%    in the investigation of white dwarf stars with gravitational red shift
%    of the order $z=\Delta \phi = 5\times 10^{-5}$. But unless we imagine that
%    the white dwarf somehow surrounds itself by a phase number 2 this
%    white dwarf study might not really test our picture.
  
  In our model the proton mass varies on the per mille level ($\Delta V \approx 2 MeV$) from phase to phase. So, of course, we would predict that $\mu$  
 should vary on the per mille level between one
  vacuum phase and another one, in strong conflict with the above limits.
  Our only explanation could be: {\em There are very few molecular clouds in the voids, so they were not measured.}
  %\end{frame}

%\begin{frame}
  
%\end{frame}

  \section{Our picture of domain walls pushed around by nucleon pressure}
 In our dark matter model there is an interaction between nucleons and
  the domain wall, 
  in as far as the nucleons have a lower potential of the order of a few MeV
  in the ``inside-dark-matter'' vacuum phase. This means that big bubbles
  of this "inside'' type will tend to be blown further up in size by the nucleons crossing the wall from the ``outside''-side and coming into the
  ``inside''-side with dramatically higher velocity. The consequent recoil of the domain wall will cause the "inside'' vacuum to become even bigger. 
  % thereby pushing as recoil
  %the wall surrounding an ``inside'' to become even bigger. 
  In fact a balance may be achieved
  in which a temperature difference between the two sides counteracts
  the pressure connected with the curvature of the wall. That should mean the
  temperature should be higher in the ``inside the dark matter'' vacuum, and
  thus the formation of stars and galaxies in the present epoch would be made more difficult. In fact very hot gas, which rather becomes plasma, is basically
  invisible, since the interaction of light with just nuclei and electrons
  corresponds to Compton scattering with a very broad spectrum and does not give a very characteristic signal.

  Domain wall universes have been studied \cite{Dwdu, Cvetic, onescale}, but we
  shall not go much into that in the present article.

  \section{After James Webb Observation Addendum}
  
  Recently the James Webb telescope discovered galaxies that had been produced 
  much earlier than expected in the development of the Universe \cite{Naidu, Adams, Webb, Labbe}, even some within the first half a milliard years after the Big Bang. We should like to present a possible explanation for such surprisingly early stars and galaxies in terms of the model discussed above with its domain walls and picture of dark matter.   

  Let us therefore give a very short survey of how our model would influence
  or not influence the early cosmology:

  \begin{itemize}
  \item From the very beginning we should expect our domain walls to be
    present due to thermal fluctuations to the degree that they make sense at
    all. In detail we have in mind as our best guess that our domain walls
    are the result of QCD with the Nambu JonaLasinio spontaneous break
    down of chiral symmetry being in more than one phase. It could be
    said that QCD in another phase was not quite in the usual confinement
    phase but that the color symmetry was rather say Higgsed -
    spontaneously broken - so as to be aligned along the breaking direction
    of the Gellman $SU(3)$, if such an alignment could make any sense.
    At least one has two $SU(3)$ groups at hand. But whatever the phase
    distinction might be, we speculate from the hoped for order of magnitude of the domain wall tension that it is involved with physics at the
    energy scale of MeV's or up to hundreds of MeV. Then it would mean
    that for higher temperatures than that, we would have the domain walls
    all over the Universe as a result of thermal fluctuations. This could give an extra, otherwise unexpected, contribution to the effective cosmological constant in say the first second after the Big Bang.

  \item As temperature falls the domain walls, starting from being distributed all over the Universe, 
  begin to contract and two types of region occur corresponding to the two
  vacua. Now the type of vacuum ending up as the inside the dark matter
  pearls one is supposed to attract the nucleons, which thus collect in
  that phase. For reasons which we would have to calculate or just speculate about
  the phase going to be the one outside the dark matter pearls wins and
  becomes at first the dominant phase. Because of the higher pressure inside the dark matter pearls or proto-pearls, the Big Bang Nucleosynthesis is
  supposed to take place earlier inside these proto-pearls.
  We have previously speculated \cite{Dark1, Dark2} that an explosive
  fusion of helium into say carbon occurs inside the pearls, where almost all the
  nucleons are collected. Hereby a fraction of the nuclei is emitted and
  becomes the ordinary matter, while the remaining pearls when contracted
  make up the dark matter.

  \item  A very few bubbles are supposed accidentally to be exceptionally
  large, and thus the pressure needed to keep them from collapsing is
  appreciably smaller. Now, because of the attraction of the nucleons, the nucleons are effectively pulled into such a large bubble making it
  even larger. The nucleons hitting the domain wall from the usual vacuum side will namely be pulled in with a force from the potential difference
  for nucleons on the two sides. This will give a reaction force pulling
  the domain wall from the inside dark matter vacuum towards the usual vacuum.
  Once the bubble has got sufficiently large the oppositely pointing force
  due to the tension in the domain wall will be so weak that the bubble
  of inside the dark matter phase will grow and grow. Also when the wall
  hits the small pearls of the inside dark matter vacuum they will unite
  with them, and the material in the hit pearls will be spit into the
  big bubble. So this will further contribute to making the very big bubble even
  larger. In this way some very few bubbles passing a critical limit can be
  very large. It would be such exceptionally large bubbles, which have
  passed the critical limit beyond which they grow instead of contracting,
  that could be the ones filling say the big voids.

  \item If the big bubbles are produced in the just described way, by growing
  up from pearls only a bit larger than the usual dark matter pearls, they
  will say in the era of Big Bang Nucleosynthesis for the ordinary matter
  not yet be so dramatically large. Thus, since there are very
  few of them, they would probably have very little significance in this
  Big Bang Nucleosynthesis Era, and we should not see much effect of them
  in the fitting of the abundances of the light elements like helium, lithium
  and the hydrogen isotopes. It is of course important for the viability
  of our model that it does not make much change in the Big Bang Nucleosynthesis, which works very well without any extra ordinary matter.
  It is rather crucial that our dark matter model, which genuinely consists
  of ordinary matter, has it so strongly packed into the pearls that the
  packed in part cannot make itself felt in the Big Bang Nucleosynthesis.

    \item But as time develops and the exceptionally large bubbles grow and grow
    they can come to cover significant parts of the space in the Universe.
    Now you then have two different types of region in the Universe:
    \begin{itemize}
    \item There is the region with the usual vacuum, which is the outside the dark
    pearls one, covered with a swarm of dark matter pearls which are
    really spheres of the other type of vacuum filled with highly compressed
    ordinary matter inside such as say carbon. Interpreting the highly pumped up
    pearls as dark matter, this part is exactly the usual picture of the
    cosmological content.
    \item There are also supposedly comparable amounts of space with the vacuum of the inside the dark matter pearl type. In these regions there can be no
    dark matter. In a formal sense such a whole huge region is a piece of
    dark matter. However this is only formally so, because the density of ordinary matter inside is only about 6 times the density of ordinary matter in the
    other type of region. But now there cannot be any true dark matter in this
    region. One could
    %If there should be any other phase in small spheres one could
    think of having small pearls of the outside the dark pearls type, but that
    does not attract the nucleons, it rather repels them. So one cannot
    make heavy pearls from that inside the dark matter vacuum. It would just
    collapse, unless one could find something else with which to pump them up.
    
    But in these regions you would have rather closely the same amount of the
    ordinary matter as one has ordinary plus dark matter in the other regions. This would mean that the formation of the clumps or fluctuations
    would be very much the same for both types of region, as long as it is
    only the question of their non-relativistic nature that counts.
\end{itemize}
    \item However if at some point the interaction with e.g. photons
    becomes important, then the two regions above could behave very differently.
    The dark matter pearls should be so
    heavy compared to the number of electrons they would likely take up
    or lose that, even if they were hot, they would not be able to emit
    so much light as the atoms that can form from the nuclei like hydrogen
    and helium. When a clump of either dark matter or ordinary matter
    contracts, it heats up, but then comes the difference: The dark matter
    interacts only very little with light and, in first approximation can not cool
    down but rather stops the contraction by the increased pressure due to the
    heating. However, for the ordinary matter which contains atoms that can emit light and get cooled, the contraction can go further heating the
    ordinary matter but it can continue to cool. So the ordinary matter
    contracts by the help of the cooling by light emission, while the dark
    matter cools much less.This leads to the picture we know:
    A big extended cloud of dark matter with a very slow variation
    of the density in space, and a much smaller clump of ordinary matter
    in the middle of the dark clump.
    
    But in a region in which the vacuum is of the inside the dark
    pearls type we have only ordinary matter - and about 6 times the density
    of the other region - but no genuine dark matter (although formally the
    whole region is a dark matter bubble). Now it can all contract as fast
    as ordinary matter by emission of light. So a clump of a certain
    geometrical size would in such a region with high ordinary matter density
    contract without leaving a much larger dark matter halo around.
    A size that should have given a small galaxy could when it is 6 times
    as dense give an appreciably more massive galaxy.
    
    This was the surprise from the James Webb early galaxy observations:
    There were at an early time more massive galaxies - with mass as large as the 
    Milky Way say - than one should expect at the time 500 million to
    700 million years after the Big Bang \cite{Labbe}. It should, according to the theoretical expectation, have been the dark age with essentially no stars and thus only rather little light.

\end{itemize}

 If this explanation of ours for the too early star and galaxy formation
should be the right one, then from the above description you may notice that
we had two regions, and the early stars and galaxies only formed in one of the
regions. In the region with the vacuum which is outside the dark pearls
the prediction of the astronomers that there should be a dark age should
hold just as predicted. So we should expect that the mysterious too early
galaxies should clump into some regions, and not be present outside these regions.

Roughly you could say one type of region starts earlier than the other type.

\subsection{Compact Galaxies}

 For the six galaxies reported in reference \cite{Labbe} to appear old and massive means they were forming
hundreds of stars a year shortly after the Big Bang. In comparison, the
Milky Way only forms around one to two new stars every year. Further,
these potential galaxies are said to be about 30 times more compact in size than
ours despite having as many stars.  ``Compact'' must mean geometrically of less extent than the Milky Way.

If it were true that they were formed in a vacuum of the type which is usually
inside the dark matter, there would  be about 6 times as much ordinary matter in the region of the recently found very early galaxies than around normal
galaxies. Furthermore, contrary to galaxies in our region of the Universe, their contraction would not be provided by dark matter but rather by ordinary matter. Since ordinary matter tends to contract much more than the dark matter, this would mean that these early galaxies should be expected to be geometrically smaller
than usual galaxies. So it would be an immediate qualitative success of our
two vacuum phase story that the surprising early galaxies are indeed more
compact.

  \section{Chance of getting rid of the Dark Energy!}
   We should call attention to the fact that the idea of having extended domain walls
  of the unexpectedly low order of magnitude of the tension, which we propose,
  gives a chance of setting the cosmological constant to zero in fits to the cosmological data, in spite of the
  famous increasing speed of Hubble expansion. This idea is contained in the work of
  Richard A. Battaye et al \cite{Dwdu}. The point is that replacing
  the energy density from the cosmological constant (the dark energy) by the
  average energy density from a system of domain walls, spread relatively
  smoothly in space on a large scale, will give 2/3 of the negative pressure
  of the corresponding energy of a cosmological constant. This means that
  if one can squeeze into the fit just 3/2 times as much domain wall energy
  as is in the present fit assigned to the cosmological constant, then one
  can get the domain walls do the job of explaining the increasing Hubble expansion in much the same way as the cosmological constant. 
  We would say that replacing the cosmological constant by domain walls is an attractive idea, if possible, although we must admit that it is a great mystery why the cosmological constant is so small
  as it is and that it would still be a mystery or perhaps even a bigger one,
  if it were even smaller. 
  %But one could also say that there must be a mechanism
  %making it small, and this would presumably end up with a cosmological
  %constant value with a random logarithm, so that at a point where one can
  %measure it to a certain accuracy it would most likely that it would
  %either look much smaller , i.e. effectively zero, or much bigger. Thinking
  %this statistics in logarithm way it would be slight advantage to be allowed
  %to take the cosmological constant - at present - to be zero. So replacing
  %it if possible by domain walls would, we would say, be an attractive
  %idea, if possible.
  
  In section \ref{section 3} we obtained an upper limit of $S^{1/3} < 100$ MeV for the tension of a single domain wall having a dimension of the visible Universe length scale. Here we will estimate the tension required if the domain walls lie along the borders of the voids and their energy density could replace the dark energy density.
  
 We take the voids to have sizes of the order of 100 million
 light years, so that the typical distance between domain walls is
 \begin{eqnarray}
 	% \hbox{taken: } 
 	\hbox{``distance between walls''}&=&(30 \; to\;  300)\, \hbox{million light years}\\
 	  &\approx& 100\;  \hbox{million light years}.
 \end{eqnarray}
  In fact we can crudely ignore whether the dark energy is 75\% or
  100\% of the energy density of the Universe, as we are only making order of
  magnitude estimates. So let us take the energy density of the dark energy, which we want to replace by domain walls in our cosmological model, to be
  \begin{eqnarray}
    \hbox{``energy density''} &=& 10^{-26}kg/m^3\\
    \hbox{and }\quad \hbox{``distance between domain walls''} &=& 10^8\, \hbox{light  \; years}
    = 10^{24} m.
\end{eqnarray}
    Then 
\begin{eqnarray}    
    \hbox{``wall surface energy density needed''} &=& 10^{24}m*10^{-26}kg/m^3
    = 10^{-2}kg/m^2\\
   % \hbox{With $c=1=\hbar$ :} 
   &=& (30\, MeV)^3.
    \end{eqnarray}
Thus the domain wall tension or energy per unit area needed for replacing dark energy has the cubic root:
\begin{eqnarray}
   S^{1/3}&=& 30\, MeV. \label{tension}
\end{eqnarray}
%\section{Conclusion}
%\begin{frame}

In our model of dark matter bubbles of radius R, the pressure provided by the surface tension S of the surrounding domain wall has to balance the pressure due to the highly relativistic degenerate electron gas inside the bubble having a 
Fermi momentum $p_f$:
\begin{equation}
	\frac{2S}{R} = \frac{1}{12\pi^2}p_f^4.
\end{equation}
Assuming there are roughly equal amounts of protons and neutrons, there must be about twice as many nucleons as electrons inside the bubble. Then it follows \cite{Tunguska} that
\begin{equation}
	S^{1/3} = \left (\frac{M}{24\pi^5m_N} \right )^{1/9}\frac{p_f}{2}, \label{S13}
\end{equation}  
where $M$ is the mass of the dark matter particle and $m_N$ is the nucleon mass. The Fermi momentum is given in terms of the change in potential energy $\Delta V$ of a nucleon when it passes through the domain wall by
\begin{equation}
	p_f = \frac{2\Delta V}{\xi^{1/4}},
\end{equation}
where $\xi = R/R_{crit}$ and $R_{crit}$ is the "critical size'' of the bubble below which it would collapse. 

From our fits \cite{Bled20} to the frequency and intensity of the 3.5 keV X-ray line observed from galaxy clusters, Andromeda and the Milky Way Center, we found that the parameter $\frac{\Delta V}{\xi^{1/4}} \approx 2\, MeV$. So equation \ref{S13} can be written in the form:
\begin{equation}
	 \frac{M}{1.24*10^{-23}kg} = \left (\frac{S^{1/3}}{2MeV}\right )^9.
\end{equation}
Then using the tension $S^{1/3} = 30\, MeV$ from equation \ref{tension} for domain walls that can replace the dark energy, we obtain the the mass of the corresponding dark matter pearl to be
\begin{eqnarray}
	 M &=& 15^9* 1.24 *10^{-23}kg\\
	&=& 4.8*10^{-13}kg.
\end{eqnarray} 
The radius of such a dark matter pearl is given by
\begin{eqnarray}
	 R&=& \frac{1}{p_f}\left ( \frac{9\pi M}{8m_N}\right )^{1/3}\\
	  &=& (4\,MeV)^{-1}*(10^{15})^{1/3}\\
	%  &=&1/4 *10^5 MeV^{-1}\\
	  %	&=& 1/4*10^5*0.2*10^{-12}m\\
	  &=& 5*10^{-9}m.
\end{eqnarray}
This size is an order of magnitude smaller than the size $10^{-7}m$ of a typical intergalactic dust grain, so that our considerations that the
dust grain typically will be larger than the bubble with the radius $R$
remains true.

%should think that
%the Hubble expansion had - on average - been 1.52 times larger. It is the
%calculation written as CMB Planck that would have become 
\section{Conclusion}
%\begin{frame}

%{\bf Conclusion}
We have worked for a long time on a model for dark matter entirely within the Standard Model, speculating that there is more than one vacuum phase and that the different phases have the same energy density. Our dark matter particles have a size of order a few nanometre and a surface tension of roughly $S = (10\, MeV)^3$. In this talk we have mainly considered possible cosmological implications of our model and we list them below.
\begin{itemize}
	\item Most likely the different vacuum-phases should have similar order
	of magnitudes of space-time extension. (Microcanonical ensemble).
	\item Different vacuum-phases have different ``constants'' of nature, e.g.
	the fine structure constant $\alpha$. Such variation has possibly been observed, but not
	yet tested for void versus cluster dependence.
		\item A similar search for variations from phase to phase of the proton to nucleon mass ratio turns out rather catastrophically for our model, in as far
	as the observations show that the deviation if any is less than of the
	order of one in ten million, while we would expect 
	deviations at the per mille level for this ratio.
	\item The very recent observations by the James Webb telescope of very early
	galaxies may be considered a success for our model, since they could appear
	in the regions with our ``inside the dark matter phase vacuum'' where
	the dark matter would have been replaced by ordinary matter.  Stars and galaxies would most likely form earlier in these regions than in regions with the usual mixture of dark and ordinary matter.
	\item The lack of galaxies today in the regions of the inside
	the dark matter phase, identified with the voids, might find an explanation by the heating up by nuclei crossing the domain wall from the outside the dark matter phase into this phase as it expands into ``our phase''
	(= the one we live in).
	\item We have proposed a physical idea in QCD with spontaneous
	breaking of chiral symmetry: Once you have a mass for the quarks,
	you can imagine that
	\begin{itemize}
		\item Either the breaking is genuinely broken by the quark mass, so
		that there is no genuine spontaneous breaking, or
		\item the quark mass is only guiding the direction of breaking in a
		genuine spontaneously broken chiral symmetry.
	\end{itemize}
	We propose a phase transition between such phases.
		With several quark
	masses a breaking in several steps may be possible.

\end{itemize}
% \end{frame}
%{\bf Conclusion}

\end{document}